\documentclass[11pt,preprintnumbers,aps,amssymb,nofootinbib,amsmath,superscriptaddress]
{revtex4}
\usepackage{epsfig,epsf,color}
\usepackage{bm} 
\usepackage{epstopdf}
\newcommand{\beq}{\begin{equation}}
\newcommand{\beql}[1]{\begin{equation}\label{#1}}
\newcommand{\eeq}{\end{equation}}
\newcommand{\eq}[1]{(\ref{#1})}
\newcommand{\fig}[1]{Fig.~\ref{#1}}
\renewcommand{\sec}[1]{Sec.~\ref{#1}}
\newcounter{topiccounter}
\renewcommand{\b}[1]{{\bm #1}} 

\newcommand{\as}{\alpha_s}

\newcommand{\aver}[1]{\left\langle #1 \right\rangle}
\newcommand{\jpsi}{J\mskip -2mu/\mskip -0.5mu\psi}

\begin{document}


\title{Nuclear modification of the $\jpsi$ transverse momentum distributions\\ in high energy $pA$ and $AA$ collisions}

\author{D.E.~Kharzeev}
\affiliation{Department of Physics and Astronomy, Stony Brook University, Stony Brook, NY 11794, USA} 
\affiliation{Department of Physics, Brookhaven National Laboratory,
Upton, NY 11973-5000, USA}

\author{E.M.~Levin}
\affiliation{HEP Department, School of Physics,
Raymond and Beverly Sackler Faculty of Exact Science,
Tel Aviv University, Tel Aviv 69978, Israel}
\affiliation{Departamento de F\'\i sica,
Universidad T$\acute{e}$cnica Federico Santa Mar\'\i a   and
Centro Cient\'\i fico-Tecnol$\acute{o}$gico de Valpara\'\i so,
Casilla 110-V,  Valparaiso, Chile}

\author{K.~Tuchin}
\affiliation{Department of Physics and Astronomy, Iowa State University, Ames, IA 50011, USA}

\date{\today}


\begin{abstract}
We evaluate the transverse momentum spectrum of $\jpsi$ (up to semi-hard momenta) in $pA$ and $AA$ collisions taking into account only the initial state effects, but resumming them to all orders in $\as^2 A^{1/3}$. In our previous papers we noticed that cold nuclear matter effects alone could not explain the experimental data on rapidity and centrality dependencies of the $\jpsi$ yield in $AA$ collisions indicating the existence of an additional suppression mechanism. Our present calculations indicate that the discrepancy persists and even increases at semi-hard transverse momenta, implying a significant final state effect on $\jpsi$ production in this kinematical domain. The QCD dipole model we employ is only marginally applicable for $\jpsi$ production at mid-rapidity at RHIC energies 
but its use is justified in the forward rapidity region. At LHC energies we can quantitatively evaluate  
 the magnitude of cold nuclear matter effects in the entire kinematical region of interest. We present our calculations of $\jpsi$ transverse momentum spectra in $pA$ and $AA$ collisions at LHC and RHIC energies. 

\end{abstract}

\maketitle

\section{Introduction}\label{sec:intr}

Understanding the nuclear modification of $\jpsi$ production is a long-standing problem of high energy nuclear physics.  The $\jpsi $ suppression was proposed as a litmus test of the quark-gluon plasma (QGP) formation \cite{Matsui:1986dk} in the early days of relativistic heavy ion program. Since then it was realized that even though $\jpsi$ 
does offer a very valuable insight into the properties of the medium, the interpretation of 
the ``anomalous" $\jpsi$ suppression is not straightforward. It is still not clear whether $\jpsi$ 
``melts" at the critical temperature \cite{Mocsy:2007yj,Mocsy:2007jz}. Moreover, $\jpsi$ production is affected by the ``cold nuclear matter" (CNM) effects that tend to suppress it in a way that may resemble the suppression in QGP. The nature of these  CNM effects is a subject of a controversy; however it is clear that the coherence in the longitudinal direction must play a pivotal role.  Since the knowledge of the CNM effects is crucial for quantifying the impact of the QGP on $\jpsi$ production, we have been motivated to perform the investigation of the  CNM effects in a series of papers \cite{Kharzeev:2005zr,Kharzeev:2008cv,Kharzeev:2008nw,Dominguez:2011cy}. There we calculated the total cross section for production of $\jpsi$ in $pA$ and $AA$ collisions in the framework of the QCD dipole model \cite{dip}.  Let us briefly explain the essence of this approach to $\jpsi$ production. At high energies, the cross section is dominated by the $t$-channel gluon exchanges. The properties of the $\jpsi$ wave function with respect to $C$ and $P$ transformations require that only an odd number of gluons are attached to the $c$ and $\bar c$ pair making up the  $\jpsi$. Taking into account the large coherence length, one can realize that the dominant contribution to the $\jpsi$ cross section comes from multiple gluon exchanges along its path through the nucleus. In a large nucleus such that $\as^2A^{1/3}\sim 1$ the gluon exchanges with different nucleons are parametrically enhanced \cite{MV,Kovchegov:1999yj,Kovchegov:1996ty}.

The study of the total cross section as a function of rapidity and centrality in  \cite{Dominguez:2011cy} indicated that the mechanism of the nuclear modification in CNM that we proposed could explain the experimental data on $\jpsi$ production in $dA$ collisions at the RHIC. In $AA$ collisions we found that it was responsible for a significant suppression of $\jpsi$ production, although it could not account for the entire effect. The total cross section is dominated by $\jpsi$'s produced at low transverse momentum due to the rapid fall-off of the differential cross section. A more detailed information is available through the study of the $\jpsi$ transverse momentum spectra that we perform in the present work.  To understand the general form of the $\jpsi$ spectrum at semi-hard momenta and at high energies note that if the saturation momentum $Q_s$ is much larger than the quark mass, the geometric scaling \cite{Stasto:2000er,Levin:1999mw,Levin:2000mv,Levin:2001cv,Iancu:2002tr,KLM} ensures that the cross section at a given centrality scales as $d\sigma/d^2p_\bot d^2b_\bot\sim Q_s^4/p_\bot^6$. The factor $(Q_s^2)^2$ indicates that at least two nucleons take part in the scattering. At high $p_\bot$ the coherence is lost, the geometric scaling breaks down and the power law of the $p_\bot$ dependence is determined by the relevant non-perturbative matrix elements instead of $Q_s^2$. The transition from the semi-hard to the  hard regime occurs at transverse momenta of the order of $Q_s^4/m^2$\cite{Iancu:2002tr,Levin:2000mv,Levin:2001cv,KLM}, but the detailed theoretical and phenomenological information about this region is scarce.  The dipole model is applicable in the semi-hard region that we investigate in this paper. 

The paper is structured as follows. In \sec{sec:x-sect} we derive the formulae for the $\jpsi$ transverse momentum spectrum in the semi-hard region by summing up all multiple interactions of the $c\bar c$ dipole with the nucleus compatible with the $\jpsi$ quantum numbers; we follow our analysis in \cite{Dominguez:2011cy}. Since we are not interested in the spectrum per se, but rather in its modification in the CNM, we re-write the cross sections for $pA$ and $AA$ collisions as a convolution of the cross section in $pp$ collisions and the corresponding nuclear-dependent scattering factors. The main result is given by \eq{xsec2},\eq{TT},\eq{TAA},\eq{F}. In \sec{sec:evol} we discuss the quasi-classical approximation and generalize our results to include the low-$x$ evolution. In particular we found that the peculiar dependence of the scattering amplitude on the longitudinal coordinate makes it possible to derive a simple approximate formula for inclusive cross section given by \eq{xsec7},\eq{sec9}. In \sec{sec:num} we perform the numerical calculations of the nuclear modification factor (NMF) using the DHJ model \cite{Dumitru:2005kb} for the forward dipole--nucleus scattering amplitude. Our results are displayed in \fig{raa0}, \fig{raaF} and \fig{raa-lhc}. Our calculation is in a reasonable agreement with  the experimental data on $dAu$ collisions at the RHIC. However, it tends to overestimate the NMF in $AuAu$ collisions, especially at high $p_T$. This probably indicates that the final state effects responsible for $\jpsi$ suppression grow at higher $p_T$. One possible scenario for an additional nuclear suppression is due to the absorption in the quark-gluon plasma. An additional contribution to the nuclear suppression was recently investigated by one of us in \cite{Marasinghe:2011bt,Tuchin:2011cg} where it was found that a strong magnetic field \cite{mag1,mag2} created in heavy ion collisions strongly suppresses high $p_T$ $\jpsi$'s. It must be stressed that the dipole model can only be used for a qualitative estimate at mid-rapidity at RHIC energies because of the short coherence length; this is illustrated in \fig{fig:long-ff} where we show the nuclear longitudinal formfactor. Neverthless, since the coherence length grows exponentially with rapidity, already at  rapidity $y=1.7$  and $p_T\lesssim 2$~GeV at RHIC the dipole model provides a fair estimate of the NMF. At the LHC the dipole model is applicable even at higher $p_T$'s. Therefore, a definitive test of the predicted enhancement of the final state effects at higher $p_T$ can be done at the LHC. To this end, we present in \fig{raa-lhc} our calculation for the center-of-mass energy 7~TeV for $pA$ and $PbPb$ collisions. This is our prediction for the CNM effect at the LHC.

\section{$\jpsi$ production cross section}\label{sec:x-sect}

The elementary process contributing to $\jpsi$ production cross section in pA collisions in the target rest frame is the scattering of a gluon from the projectile proton wave function on the target nucleus. In $gA$ collisions the $c\bar{c}$ pair emerging from the gluon splitting is in the adjoint color representation. The $c\bar{c}$ forming the $J/\psi$ is a color singlet. As it has been noted in \cite{Dominguez:2011cy}, in the large-$N_c$ approximation there is a particular dipole-nucleon inelastic collision, at the longitudinal coordinate $\xi$, which converts the adjoint $c\bar{c}$ pair to the color singlet one. Later interactions, occurring after the $c\bar{c}$ pair is in a singlet state, are purely elastic and keep the singlet intact. Earlier interactions, occurring while the $c\bar{c}$ is in the adjoint representation, may be either elastic, occurring off a \emph{single} $c$ or $\bar{c}$ in the amplitude or complex conjugate amplitude, or inelastic involving the $c$ or $\bar{c}$ in both the amplitude and complex conjugate amplitude. We therefore calculate separately the contributions to the cross section that occur before and after the last inelastic interaction and the inelastic interaction itself.

\subsection{Multiple scattering before and after the last inelastic interaction}
\begin{figure}[ht]
      \includegraphics[height=6cm]{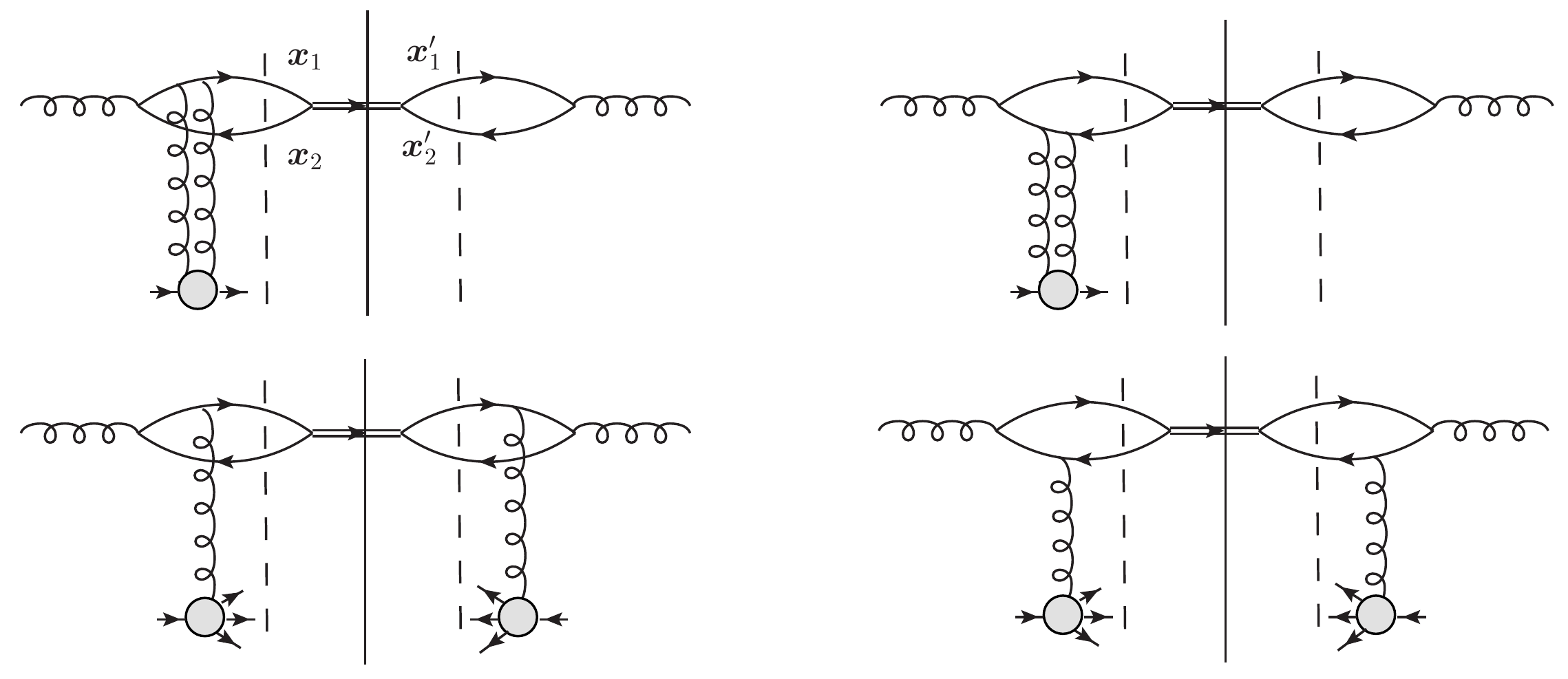} 
  \caption{One of the interactions before the last inelastic scattering. The diagrams that are complex conjugate to the first row of diagrams are not shown. The vertical dashed line denotes the last inelastic interaction when the $c\bar c$ pair is converted into the color-singlet state. The vertical solid line is the cut corresponding to the final state.}
\label{fig.before}
\end{figure}

The sum of diagrams depicted on \fig{fig.before} gives the contribution of elastic and  inelastic scatterings of color octet before the last inelastic interaction.  We have
\beql{one.before}
\frac{1}{2}\left( V(\b x_1-\b x_1')-V(0)\right)+\frac{1}{2}\left( V(\b x_2-\b x_2')- V(0)\right)\,,
\eeq
where 
\beql{g1}
V(\b x) = \int d^2q\, e^{-i\b q\cdot \b x}\frac{1}{\sigma_{gN}}\frac{d\sigma_{gN}}{d^2q}\,,
\eeq
is the Fourier transformation of the normalized gluon-nucleon cross section.
 Let us introduce notations
\begin{align}
&\b x_1 = \b b+\frac{1}{2}\b r\,,\quad \b x_2 = \b b-\frac{1}{2}\b r\,,\\
& \b x_1' = \b b'+\frac{1}{2}\b r'\,,\quad \b x_2' = \b b'-\frac{1}{2}\b r'\,,\\
& \b \Delta = \b b -\b b'\,, \quad \b B = \frac{1}{2}(\b b+\b b')\,.
\end{align}
Multiplying \eq{one.before} by $\rho T(B) \sigma$, where $\rho$ is the nuclear density, $\sigma$ is the total dipole-nucleon cross section, $T(B)$ is the nuclear thickness,  we get 
\beql{1b0}
-\frac{1}{8}Q_s^2(B) \left[ (\b x_1-\b x_1')^2\ln\frac{1}{\mu|\b x_1-\b x_1'|}+ (\b x_2-\b x_2')^2\ln\frac{1}{\mu|\b x_2-\b x_2'|} \right]\,,
\eeq
where $Q_s^2$ is the gluon saturation scale given by \eq{sat} and $\mu$ an infrared cutoff.  Eq.~\eq{1b0} is the lowest order expansion of the dipole-nucleus scattering amplitude. It provides the initial condition for the low-$x$ evolution \cite{Balitsky:1995ub,Kovchegov:1999yj}. This evolution erases the dependence of the scattering amplitude on the infrared scale $\mu$ so that the amplitude becomes dependent only on the saturation scale -- the effect known as the geometric scaling \cite{Stasto:2000er,Levin:1999mw,Levin:2000mv,Levin:2001cv,Iancu:2002tr}. This allows to drop the logarithmic factors, as suggested by  Golec-Biernat and Wusthoff \cite{MOD},  and to write 
\beql{1b}
-\frac{1}{8}Q_s^2(B) \left[ (\b x_1-\b x_1')^2+ (\b x_2-\b x_2')^2 \right]= -\frac{1}{4}Q_s^2(B)\left( \Delta^2+\frac{1}{4}(\b r-\b r')^2\right)\,.
\eeq
Our notation is $\b x^2 = x^2= x_\bot^2$.

\begin{figure}[ht]
      \includegraphics[height=6cm]{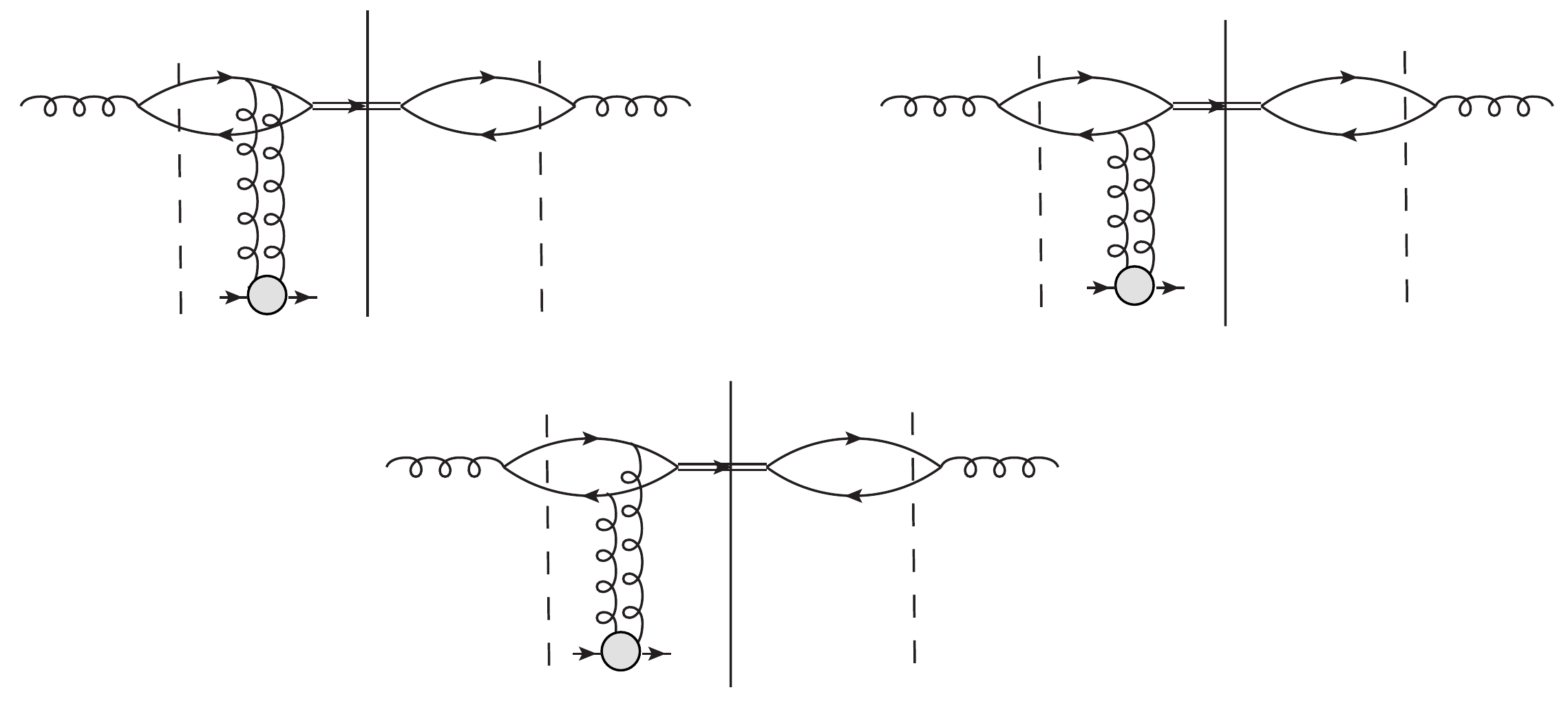} 
  \caption{One of the interactions after the last inelastic interactions. Complex conjugate diagrams  are not shown.}
\label{fig.after}
\end{figure}
Sum of the diagrams on \fig{fig.after} represents the contribution of an elastic scattering of color singlet after the last inelastic scattering. The corresponding factor is given by 
\beql{1a}
-\frac{1}{8}Q_s^2(B) \left[(\b x_1-\b x_2)^2+(\b x_1'-\b x_2')^2\right] = -\frac{1}{8}Q_s^2(B)( r^2+ r'^2)\,,
\eeq
 where we again dropped the logarithmic factors.

\subsection{The last inelastic interaction}
\begin{figure}[ht]
      \includegraphics[height=6cm]{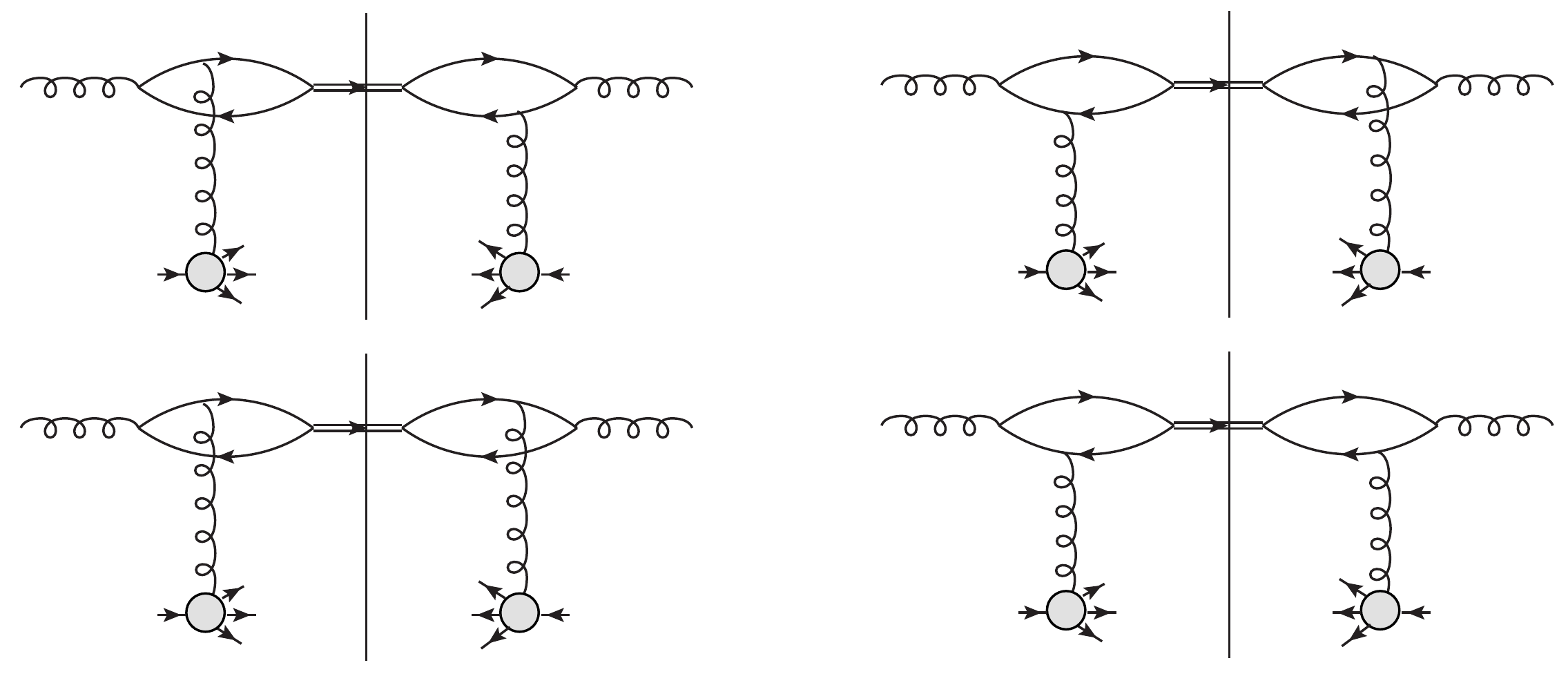} 
  \caption{The last inelastic interaction.}
\label{fig.last}
\end{figure}
Sum of the diagrams on \fig{fig.last} represents the contribution of the last inelastic scattering converting the color octet into the color singlet. Neglecting logarithms we get
\beql{1L}
-\frac{1}{8}Q_s^2(B)\left[ (\b x_2-\b x_1')^2+(\b x_1-\b x_2')^2-(\b x_1-\b x_1')^2-(\b x_2-\b x_2')^2\right] = -\frac{1}{4}Q_s^2(B)\,\b r\cdot \b r'\,.
\eeq
This formula correctly describes the behavior of the total cross section, which receives its main contribution from the small-$p_\bot$ region. However, since we are interested in the $p_\bot$-spectrum, \eq{1L} must be corrected by the logarithmic factors. At the leading order in $\as$, Eq. \eq{1L} is the only contribution to the cross section, while \eq{1b} and \eq{1a} contribute at higher orders (multiple scatterings).  Therefore, for a consistent analysis it is 
necessary to study the contribution of the diagrams of \fig{fig.last} to the total cross section.

\subsection{Logarithmic contributions to the last inelastic scattering}

The amplitude for $\jpsi$ production in a $pN$ collision can be written as \cite{Brodsky:1994kf}
 \beq
A(p_\bot)= \frac{1}{p_\bot^2}\mathcal{F}(p_\bot) \,,
\eeq
where $\mathcal{F}(p_\bot)$ is the  form-factor given by
\begin{align}\label{fff}
\mathcal{F}(p_\bot)  =& \int \frac{d^3k}{(2\pi)^3}\Psi_G( k)\,\Psi_V( k- p)= 2\pi\as \int_0^1 dz\int \frac{d^2r}{4\pi}\Phi(r,z)\left( e^{-i\frac{1}{2}\b r\cdot \b p}-e^{i\frac{1}{2}\b r\cdot \b p}\right)
\,.
\end{align}
Here $\Psi_G$ and $\Psi_V$ are the wave functions of the $s$-channel gluon and $\jpsi$ correspondingly, and $\Phi$ is explicitly given by \eq{Phiphoton}. In deriving \eq{fff} we assumed that the coherence length is much larger than the nuclear radius. This is strictly true in the Regge limit. However, at RHIC and LHC the effects of a finite coherence length may be important, especially for heavy particles and/or high transverse momenta. In Appendix~\ref{appB} we compute
the longitudinal form factor $\mathcal{F}_L$ that is a measure of the longitudinal coherence, and discuss the kinematics in which our results in this paper are applicable.  

Introduce the impact parameter representation of the scattering amplitude 
\beq\label{impact}
A(b) = \int \frac{d^2\ell}{(2\pi)^2} \,A(\ell)\, e^{-i\b b\cdot \b \ell}= \frac{\as}{2\pi}\int \frac{d^2\ell}{\ell^2}\int \frac{d^2r }{4\pi}\Phi (r) \left( e^{-i(\b b+\frac{1}{2}\b r)\cdot \b \ell}-e^{-i(\b b-\frac{1}{2}\b r)\cdot \b \ell}\right)\,.
\eeq
Then the cross section for scattering on a given nucleon reads
\beq
\frac{d\sigma_{gN}}{d^2p_\bot} =|A(p_\bot)|^2 = \int d^2 b \int d^2b'\, e^{i\b p\cdot(\b b-\b b') }A(b)A^*(b')\,,
\eeq
To obtain the cross section for scattering on a nucleus we need to average the forward scattering amplitude over all possible positions of the nucleon in the nucleus. We have
\beql{aver0}
\frac{d\sigma_{gA}}{d^2p_\bot} = \int d^2 b \int d^2b'\, e^{i\b p\cdot(\b b-\b b') }\aver{A(b)A^*(b')}\,.
\eeq
where
\beq\label{aver1}
\aver{\ldots }= \int d^2b_a\, \rho T(b_a)A(b-b_a)A^*(b'-b_a)\,.
\eeq
Here $\b b_a$ is the transverse position of the nucleon with respect to the center of the nucleus, $\rho $ is the nuclear density and $T(b)$ is the nuclear thickness. For a big nucleus $\rho T(b_a)\approx A/S_A$ is a constant, with $S_A$ being the nuclear cross-sectional area.
We have
\begin{align}
\aver{A(b)A^*(b')}=&\frac{A}{S_A}\frac{\as^2}{(2\pi)^2}\int d\Gamma\int d^2b_a  \int \frac{d^2\ell}{\ell^2}\left( e^{-i(\b b-\b b_a+\frac{1}{2}\b r)\cdot \b \ell}-e^{-i(\b b-\b b_a-\frac{1}{2}\b r)\cdot \b \ell}\right)\nonumber\\
&\times
\int \frac{d^2\ell'}{\ell'^2}\left( e^{i(\b b'-\b b_a+\frac{1}{2}\b r')\cdot \b \ell'}-e^{i(\b b'-\b b_a-\frac{1}{2}\b r')\cdot \b \ell'}\right)\\
=&\frac{A}{S_A}\frac{\as^2}{(2\pi)^2}\int d\Gamma\int\frac{d^2\ell}{\ell^4}e^{-i(\b b-\b b')\cdot \b \ell}\nonumber\\
&\times\left( e^{i\frac{1}{2}(\b r-\b r')\cdot \b \ell}+e^{-i\frac{1}{2}(\b r-\b r')\cdot \b \ell}-e^{-i\frac{1}{2}(\b r+\b r')\cdot \b \ell}-e^{i\frac{1}{2}(\b r+\b r')\cdot \b \ell}\right)\,.  \label{a-aver}
\end{align}
We introduced a convenient notation 
\beql{not-f}
\int d\Gamma = \int_0^1 dz  \int \frac{d^2r}{4\pi} \Phi (r,z)  \int_0^1 dz'  \int \frac{d^2r'}{4\pi} \Phi (r',z') 
\eeq
Substitution of \eq{a-aver} into \eq{aver0} gives
\begin{align}\label{main}
\frac{d\sigma_{gA}}{d^2p_\bot} =&\frac{A}{S_A}\frac{\as^2}{(2\pi)^2} \int d^2 b \int d^2b'\, e^{i\b p\cdot(\b b-\b b') }\int d\Gamma \nonumber\\
&\times\int\frac{d^2\ell}{\ell^4}e^{-i(\b b-\b b')\cdot \b \ell}\left( e^{i\frac{1}{2}(\b r-\b r')\cdot \b \ell}+e^{-i\frac{1}{2}(\b r-\b r')\cdot \b \ell}-e^{-i\frac{1}{2}(\b r+\b r')\cdot \b \ell}-e^{i\frac{1}{2}(\b r+\b r')\cdot \b \ell}\right).
\end{align}

If we were  interested in the total cross section, then integrating \eq{main} over $p_\bot$ would give
\begin{align}
\frac{d\sigma_{gA}}{d^2b}&=\as^2 \frac{A}{S_A}\int d\Gamma\int\frac{d^2\ell}{\ell^4}\left( e^{i\frac{1}{2}(\b r-\b r')\cdot \b \ell}+e^{-i\frac{1}{2}(\b r-\b r')\cdot \b \ell}-e^{-i\frac{1}{2}(\b r+\b r')\cdot \b \ell}-e^{i\frac{1}{2}(\b r+\b r')\cdot \b \ell}\right)\label{tot-main}\\
&\approx \as^2 \frac{A}{S_A} \int d\Gamma\int\frac{d^2\ell}{\ell^2}\, \frac{1}{2}\b r\cdot \b r'\,.
\label{tot-m-exp}
\end{align}
where we expanded the exponents in \eq{tot-main} because only small momenta  $\ell\sim 1/R_A$ give the logarithmic contribution to the integral. Note that at the leading order in $\as$, \eq{tot-m-exp} averages to zero. This is because $\jpsi$ is $C$ and $P$ odd and therefore there must be an odd number of gluons hooked to the fermion line, whereas \eq{tot-m-exp} accounts only for two gluons. However, once the multiple scatterings are taken into account, the $\b r\cdot \b r'$ term selects the terms with the required parity form the scattering factors before and after the last  inelastic interaction as we demonstrate below. Thus, we think of \eq{tot-m-exp} as the projection operator onto the $1^{--}$ state. Defining the saturation momentum
\beql{sat}
Q^2 = 4\pi \as^2 \rho T(B) 
\eeq
and dropping the $\ln(1/\mu|\b r\pm \b r'|)$ factors we cast \eq{tot-m-exp}  into the form
\beql{an-for}
\frac{d\sigma_{gA}}{d^2b}= \int d\Gamma\, Q_s^2\,\frac{1}{4}\b r\cdot \b r'\,.
\eeq
This is precisely the approximation in which our formula \eq{1L} is valid. We observe that in general, expansion \eq{tot-m-exp} cannot be performed for the differential cross section \eq{main}.  Indeed, in \eq{main} at high $p_\bot$, $\b \Delta= \b b-\b b'$ is small and therefore $\b \ell$ is large. Hence, we write \eq{main} as 
 \begin{align}\label{main2}
\frac{d\sigma_{gA}}{d^2B \,d^2p_\bot} =&\frac{A}{S_A}\frac{\as^2}{(2\pi)^2} \int d^2 \Delta \, e^{i\b p\cdot\b \Delta }\int d\Gamma\, \pi \, J(\b r, \b r', \b \Delta)\,.
\end{align}
where we introduced an auxiliary function $J$ as 
\begin{align}\label{J0}
J(\b r, \b r',\b \Delta)\equiv&\int\frac{d\ell^2}{\ell^4}e^{-i\b\Delta\cdot \b \ell}\left( e^{i\frac{1}{2}(\b r-\b r')\cdot \b \ell}+e^{-i\frac{1}{2}(\b r-\b r')\cdot \b \ell}-e^{-i\frac{1}{2}(\b r+\b r')\cdot \b \ell}-e^{i\frac{1}{2}(\b r+\b r')\cdot \b \ell}\right)\,.
\end{align}
Using formula \eq{int-4} of  Appendix~\ref{appA} we find
\begin{align}\label{J}
J(\b r, \b r',\b \Delta)&=
\frac{2\pi}{4} \left[-\left(\frac{1}{2}(\b r-\b r')-\b \Delta\right)^2\ln \frac{1}{\mu|\frac{1}{2}(\b r-\b r')-\b \Delta |}\right.
\nonumber\\
&\left. -
\left(\frac{1}{2}(\b r-\b r')+\b \Delta\right)^2\ln \frac{1}{\mu|\frac{1}{2}(\b r-\b r')+\b \Delta |}\right.
\nonumber\\
&\left.
+
\left(\frac{1}{2}(\b r+\b r')+\b \Delta\right)^2\ln \frac{1}{\mu|\frac{1}{2}(\b r+\b r')+\b \Delta |}
\right. \nonumber\\
&\left. +
\left(\frac{1}{2}(\b r+\b r')-\b \Delta\right)^2\ln \frac{1}{\mu|\frac{1}{2}(\b r+\b r')-\b \Delta |}
\right].
\end{align}
In particular, $J(\b r, \b r', \b 0)= \b r\cdot \b r'$ modulo the logarithmic factors.

\subsection{Summing up the multiple scatterings}

Suppose that the longitudinal coordinate of the last inelastic scattering is $\xi$. Then using \eq{1b},\eq{1a},\eq{main2} and summing over all possible number of interactions, we find that  the scattering amplitude for $\jpsi$ production in $pA$ collisions  is given by 
\begin{align}\label{TpA}
T_{pA\to \jpsi X}(\b r,\b r',\b B,\b \Delta)=&\int_0^{T(B)} \frac{1}{4}\frac{Q_s^2 }{T}J(\b r, \b r',\b \Delta) \, e^{-\frac{1}{4}Q_s^2\left[\Delta^2+\frac{1}{4}(\b r-\b r')^2\right]\frac{\xi}{T}}\, e^{-\frac{1}{8}Q_s^2 (r^2+r'^2)\left(1-\frac{\xi}{T}\right)}\nonumber\\
=& \frac{4J(\b r, \b r',\b \Delta)}{(\b r+\b r')^2-4\Delta^2}\left\{ e^{-\frac{1}{16}Q_s^2(B)[(\b r-\b r')^2+4\Delta^2 ]} - 
e^{-\frac{1}{8}Q_s^2(B)(r^2+r'^2)}  \right\}\,.
\end{align}
The cross section for inclusive $\jpsi$ production in  $pA$ collisions reads
\begin{align}\label{xsec1}
\frac{d\sigma_{pA\to \jpsi X}}{d^2p_\bot dy d^2B}= & x_1G(x_1,m_c^2)\int_0^1 dz\int \frac{d^2r}{4\pi}\Phi(r,z)\int_0^1 dz' \int \frac{d^2r'}{4\pi}\Phi(r',z')\nonumber\\
&\times \int \frac{d^2\Delta}{(2\pi)^2}e^{i \b\Delta\cdot \b p}\, T_{pA\to \jpsi X}(\b r, \b r', \b \Delta, \b B)
\end{align}
where
\beq\label{Phiphoton}
\Phi(\bm r,z)=\frac{g}{\pi\sqrt{2N_c}}\left\{m_c^2K_0(m_cr)\phi_T(r,z)-\left[z^2+(1-z)^2\right]m_cK_1(m_cr)\partial_r\phi_T(r,z)\right\}
\eeq
with 
\begin{equation}\label{phiT}
\phi_T(r,z)=N_Tz(1-z)\exp\left[-\frac{r^2}{2R_T^2}\right]
\end{equation}
and where $N_T=1.23$, $R_T^2=6.5$ GeV$^{-2}$. Integrating over $p_\bot$ we reproduce the result of \cite{Dominguez:2011cy}. The color factor in \eq{Phiphoton} includes projection onto the color singlet state \cite{Dominguez:2011cy}.

It is well-known that the leading order calculation does not properly describe the transverse momentum distribution of hadrons. Moreover, the high-$p_T$ tail of such distribution stems from the short distance ``hard" processes, which violate the geometric scaling. These hard processes factorize from the rest of the scattering amplitude in a number of recently studied reactions \cite{Chirilli:2011km}. Since we are primarily interested in the nuclear modification of $\jpsi$ production in $pA$ and $AA$ collisions as compared to $pp$ ones, we propose a model for the function $J$ of \eq{J0}, which retains the essential features of  \eq{J} and additionally describes the $\jpsi$ spectrum in $pp$ collisions.  In view of \eq{tot-m-exp}  we suggest to write 
\beql{modJ} 
J(\b r, \b r', \b \Delta)\approx \b r\cdot \b r' \, F(\Delta)\,,
\eeq
where the short distance effects are described by the function $F(\Delta)$, which is  defined as
\beql{F}
F(\Delta) =\frac{1}{\sigma_{pp\to \jpsi X}} \int \frac{d\sigma_{pp\to \jpsi X}(p_\bot)}{ d^2 p_\bot} e^{-i\b p\cdot \b \Delta}d^2p_\bot 
\,,
\eeq
with $\sigma_{pp}$ being the total inelastic $pp$ cross section. The factor $\b r\cdot \b r'$ in \eq{modJ} selects (after integration over the angle between $\b r$ and $\b r'$) only  those contributions to the cross section that contain odd number of gluons hooked up to the fermion line in agreement with the $\jpsi$ quantum numbers. 

According to our proposal we model the scattering amplitude \eq{TpA} by 
\beql{TpAmod}
 T_{pA\to \jpsi X}(\b r, \b r',  B,  \Delta)=\frac{4\b r\cdot \b r'\, F(\Delta)}{(\b r+\b r')^2-4\Delta^2}\left\{ e^{-\frac{1}{16}Q_s^2(B)[(\b r-\b r')^2+4\Delta^2 ]} - 
e^{-\frac{1}{8}Q_s^2(B)(r^2+r'^2)}  \right\}\,.
\eeq
Expanding \eq{TpAmod} to the leading order in $Q_s^2$ we obtain the scattering factor for $pp\to \jpsi X$ process:
\begin{align}\label{Tpp}
 T_{pp\to \jpsi X}(\b r, \b r',  B,  \Delta)=&\frac{1}{4}\b r\cdot \b r' Q_s^2\,F(\Delta) +\frac{1}{128}\b r\cdot \b r'(-3r^2-3r'^2+2\b r\cdot \b r'-4\Delta^2)Q_s^4 F(\Delta)\,.
\end{align}
Upon averaging over the angle between the dipoles $\b r$ and $\b r'$ we get
\begin{align}\label{Tpp-a}
\aver{T_{pp\to \jpsi X}(\b r, \b r',  B,  \Delta)} = \frac{1}{64}(\b r\cdot \b r')^2 Q_s^4 F(\Delta)\,.
\end{align}
Appearance of $Q_s^4$ indicates that at least two scatterings are required to produce $\jpsi$ in accordance with the $C$ and $P$ odd nature of the $\jpsi$ wave function.  In our model $F(\Delta)$ completely determines the $p_\bot$-spectrum.\footnote{Integration with the complete function $J$ selects dipole sizes $r,r'\sim 1/p_\bot$ so that the cross section falls off as $Q_s^4/p_\bot^{6}$ at large $p_\bot$.}   This can be readily seen by substitution of \eq{Tpp-a} into \eq{xsec1} and using \eq{F}.
Our goal is to calculate the modification of the $\jpsi$ spectrum  in the cold nuclear medium. 

To generalize the scattering amplitude to $AA$ collisions, we assume that $c\bar c$ scatters independently on each nucleus.\footnote{There is no rigorous proof of such $AA$ factorization. In fact, because of  immense complexity of this problem no systematic attempt has been made to prove it. A notable exception is Ref.~\cite{Kovchegov:2000hz} that argued in favor of $AA$ factorization in single inclusive gluon production at the leading order.} The resulting amplitude reads 
\begin{align}\label{TAA}
 T_{A_1A_2\to \jpsi X}(\b r, \b r',  B,  \Delta)=&\frac{C_F}{2\as\pi^2}\frac{Q_{s1}^2 Q_{s2}^2}{Q_{s1}^2+Q_{s2}^2}\,\frac{4\b r\cdot \b r'\, F(\Delta)}{(\b r+\b r')^2-4\Delta^2}\nonumber\\
&
\times \left\{ e^{-\frac{1}{16}(Q_{s1}^2+Q_{s2}^2)[(\b r-\b r')^2+4\Delta^2 ]} - 
e^{-\frac{1}{8}(Q_{s1}^2+Q_{s2}^2)(r^2+r'^2)}  \right\}\,.
\end{align}
The corresponding cross section for inclusive $\jpsi$ production in  $AA$ collisions is given by
\begin{align}\label{xsec2}
\frac{d\sigma_{A_1A_2\to \jpsi X}}{d^2p_\bot dy d^2B_1\,d^2B_2}= &\int_0^1 dz\int \frac{d^2r}{4\pi}\Phi(r,z)\int_0^1 dz' \int \frac{d^2r'}{4\pi}\Phi(r',z')\, 2 T_{A_1A_2\to \jpsi X}(\b r, \b r',  B,p_\bot)\,,
\end{align}
where
\beql{TT}
 T_{A_1A_2\to \jpsi X}(\b r, \b r',  B,p_\bot)= \int \frac{d^2\Delta}{(2\pi)^2}e^{i \b\Delta\cdot \b p}\,  T_{A_1A_2\to \jpsi X}(\b r, \b r',  B,\Delta)
\eeq
We will simplify the nuclear profiles by the step-function. 

\section{Quasi-classical approximation and evolution effects}\label{sec:evol}

\subsection{Quasi-classical approximation}

Numerical integration over $\Delta$  in \eq{TT} is difficult as the integrand is an oscillating function.  In the quasi classical approximation we can avoid  this problem if we neglect the logarithmic dependance of the saturation momentum on the dipole size as discussed in the previous section. Let us introduce the  following Fourier representation 
\begin{align}\label{con}
&\frac{4\b r\cdot \b r'}{(\b r+\b r')^2-4\Delta^2}\left\{ e^{-\frac{1}{16}Q_s^2(B)[(\b r-\b r')^2+4\Delta^2 ]} - 
e^{-\frac{1}{8}Q_s^2(B)(r^2+r'^2)}  \right\}=
\int d^2k_\bot\,e^{-i\b \Delta\cdot\b k}\,H(\b r, \b r',  B,  \b k)\,.
\end{align}
Using \eq{TAA}, \eq{TT} and \eq{con} we cast the scattering amplitude in the form of a convolution
\beql{con3}
 T_{A_1A_2\to \jpsi X}(\b r, \b r',  B,p)= \frac{C_F}{2\as\pi^2}\frac{Q_{s1}^2 Q_{s2}^2}{Q_{s1}^2+Q_{s2}^2}\,\int F_{\b k}\,H(\b r, \b r', B, \b p-\b k)\,d^2k_\bot\,,
\eeq
where 
\beql{F-four}
F_{\b k}= \int\frac{d^2\Delta}{(2\pi)^2}e^{i\b\Delta\cdot \b k}F(\Delta)\,.
\eeq
Note, that by \eq{F} $\int d^2k  F_{\b k} = F(0)=1$.
The $\Delta$-integral in the inverse of \eq{con} can be done analytically. Let us invert \eq{con}  and use \eq{TpA} to write it  as an integral over the longitudinal coordinate $\xi'=\xi/T(B)$:
\begin{align}\label{take1}
H(\b r, \b r', B, \b k)=& \int \frac{d^2\Delta}{(2\pi)^2}\,e^{i\b \Delta\cdot\b k} \b r\cdot \b r'\,\frac{(Q_{s1}^2+Q_{s2}^2)}{16}\nonumber\\
&\times\int_0^1d\xi' 
e^{-\frac{1}{4}(Q_{s1}^2+Q_{s2}^2)(\Delta^2+\frac{1}{4}(\b r-\b r')^2)\xi'} 
\,e^{-\frac{1}{8}(Q_{s1}^2+Q_{s2}^2)(r^2+r'^2)(1-\xi')}\\
=&\int_0^1d\xi' \,\frac{\b r\cdot \b r'}{16\pi\xi'}e^{-\frac{1}{4}(Q_{s1}^2+Q_{s2}^2)\frac{1}{4}(\b r-\b r')^2\xi'} 
\,e^{-\frac{1}{8}(Q_{s1}^2+Q_{s2}^2)(r^2+r'^2)(1-\xi')}\,e^{-\frac{k_\bot^2}{(Q_{s1}^2+Q_{s2}^2)\xi'}}\label{take2}
\end{align}

Finally, the cross section reads 
\begin{align}\label{xsec3}
\frac{d\sigma_{A_1A_2\to \jpsi X}}{d^2p_\bot dy d^2B_1\,d^2B_2}= & \int_0^1 dz\int \frac{d^2r}{4\pi}\Phi(r,z)\int_0^1 dz' \int \frac{d^2r'}{4\pi}\Phi(r',z')\int d^2 k_\bot \int_0^1d\xi'\nonumber\\
&\times \frac{C_F}{\as\pi^2}\frac{Q_{s1}^2 Q_{s2}^2}{Q_{s1}^2+Q_{s2}^2}\,F_{\b k}\,\frac{\b r\cdot\b r'}{16\pi \xi'}\,e^{-\frac{p^2_\bot+k^2_\bot}{(Q_{s1}^2+Q_{s2}^2)\xi'}}\nonumber\\
&\times \,I_0\left( \frac{2p_\bot k_\bot}{(Q_{s1}^2+Q_{s2}^2)\xi'}\right) \,
e^{-\frac{1}{4}(Q_{s1}^2+Q_{s2}^2)\frac{1}{4}(\b r-\b r')^2\xi'} 
\,e^{-\frac{1}{8}(Q_{s1}^2+Q_{s2}^2)(r^2+r'^2)(1-\xi')}\,
\end{align}
where $I_0$ is a modified Bessel function and $\b B_{1,2}$ are the impact parameters of the two nuclei.

\subsection{Low-$x$ evolution}

Explicit integration over  $\b \Delta$ is not possible if the amplitude is evolved at low $x$. Nevertheless,  we found a very good approximation to our formulas even in this case. 
Let us again re-write the cross section \eq{xsec2},\eq{TAA} as an integral over the longitudinal coordinate $\xi'$ using  \eq{TpA}.
\begin{align}\label{xsec4}
\frac{d\sigma_{A_1A_2\to \jpsi X}}{d^2p_\bot dy d^2B_1d^2B_2}= &\frac{C_F}{4\as\pi^2}Q_{s1}^2 Q_{s2}^2 \int_0^1d\xi'\int\frac{d^2\Delta}{(2\pi)^2}e^{i\b\Delta\cdot \b p}F(\Delta) e^{-\frac{1}{4}(Q_{s1}^2+Q_{s2}^2)\Delta^2 \xi'}
\Xi(\xi')\,,
\end{align}
where we introduced an auxiliary function $\Xi$ as
\begin{align}\label{Xi}
\Xi (\xi') =   &\int_0^1 dz\int \frac{d^2r}{4\pi}\Phi(r,z)\int_0^1 dz' \int \frac{d^2r'}{4\pi}\Phi(r',z') \, \b r\cdot \b r'\nonumber\\
&e^{-\frac{1}{16}(Q_{s1}^2+Q_{s2}^2)(\b r-\b r')^2\xi'} 
\,e^{-\frac{1}{8}(Q_{s1}^2+Q_{s2}^2)(r^2+r'^2)(1-\xi')}\,.
\end{align}
Function $\Xi$ varies with $\xi'$ much slower than the exponential factors in \eq{xsec4}. It starts at $\Xi(0)=0$ (note the integration over the angle between $\b r$ and $\b r'$) and rises at most as a power towards a constant $\Xi(1)$. We verified this statement with a numerical calculation. We therefore approximate \eq{xsec4} as
\begin{align}\label{xsec5}
\frac{d\sigma_{A_1A_2\to \jpsi X}}{d^2p_\bot dy d^2B_1d^2B_2}\approx &\frac{C_F}{4\as\pi^2}Q_{s1}^2 Q_{s2}^2 \aver{\Xi}\int_0^1d\xi'\int\frac{d^2\Delta}{(2\pi)^2}e^{i\b\Delta\cdot \b p}F(\Delta) e^{-\frac{1}{4}(Q_{s1}^2+Q_{s2}^2)\Delta^2 \xi'}\,,\nonumber\\
&\frac{C_F}{4\as\pi^2}\frac{Q_{s1}^2 Q_{s2}^2}{Q_{s1}^2 +Q_{s2}^2} \aver{\Xi}\int\frac{d^2\Delta}{(2\pi)^2}\frac{4}{\Delta^2}e^{i\b\Delta\cdot \b p}F(\Delta) 
\left( 1-e^{-\frac{1}{4}(Q_{s1}^2+Q_{s2}^2)\Delta^2}\right)\,,
\end{align}
where the average value of $\Xi$ is given by
\beql{averxi}
\aver{\Xi} = \int_0^1 \Xi(\xi')d\xi'\,.
\eeq
The $p_\bot$-integrated cross section reads using \eq{xsec5}
\beql{xsec6}
\frac{d\sigma_{A_1A_2\to \jpsi X}}{ dy d^2B_1d^2B_2}= \frac{C_F}{4\as\pi^2}Q_{s1}^2 Q_{s2}^2 \aver{\Xi}\,.
\eeq
Now, employing \eq{xsec6} in \eq{xsec5} we obtain 
\begin{align}\label{xsec7}
\frac{d\sigma_{A_1A_2\to \jpsi X}}{d^2p_\bot dy d^2B_1d^2B_2}=\frac{d\sigma_{A_1A_2\to \jpsi X}}{ dy d^2B_1d^2B_2}
\frac{1}{Q_{s1}^2+Q_{s2}^2}\int \frac{d^2\Delta}{(2\pi)^2}e^{i\b\Delta\cdot \b p}F(\Delta) \frac{4}{\Delta^2}\left( 
1-e^{-\frac{1}{4}(Q_{s1}^2+Q_{s2}^2)\Delta^2}\right)\,.
\end{align}

We can incorporate the low-$x$ evolution effects  by the following replacement \cite{Kharzeev:2008cv,Kharzeev:2008nw,Dominguez:2011cy}
\begin{align}\label{ff1}
e^{-\frac{1}{16}Q_s^2 r^2} &\to  1-N_A(\b r/2,\b b, y)\,.
\end{align}
That the adjoined amplitude appears in the right-hand-side of \eq{ff1} is the result of the particular structure of the scattering amplitude  discussed in \sec{sec:x-sect} (see \cite{Dominguez:2011cy} for more details). Eq.~\eq{ff1} is an identity at $y=0$, whereas at $y>0$ it assumes a factorization of the two nuclei in the coordinate space for each class of diagrams  in  \sec{sec:x-sect}. Such approach to particle production in AA collisions was advocated in \cite{Kovchegov:2000hz} and has been used in most phenomenological applications. The cross section now reads
\begin{align}\label{sec9}
\frac{d\sigma_{A_1A_2\to \jpsi X}}{d^2p_\bot dy d^2B_1\,d^2B_2}=&\frac{d\sigma_{A_1A_2\to \jpsi X}}{ dy d^2b\,d^2B}
\int \frac{d^2\Delta}{(2\pi)^2}e^{i\b\Delta\cdot \b p}F(\Delta) \nonumber\\
& \times \left(  1- \left[ 1-N_A^{(1)}\left(\b \Delta,\b B_1, y\right)\right]\left[ 1-N_A^{(2)}\left(\b\Delta,\b B_2, -y\right)\right] \right)\nonumber\\
& \times \left(  1- \left[ 1-N_A^{(1)}\left(\b \Delta,\b B_1, y\right)\right]\left[ 1-N_A^{(2)}\left(\b\Delta,\b B_2, -y\right)\right] \right)^{-1}_\text{LT}\,,
\end{align}
where $\text{LT}$ stands for the ``leading twist" which is the leading term in small $\Delta$ expansion.
The result for the $p_\bot$-integrated cross section can be found  in our previous paper \cite{Dominguez:2011cy}.

The experimental data is expressed in terms of the nuclear modification factor (NMF). It is  defined as
\beql{nmf}
R_{A_1A_2}= \frac{\int_\mathcal{S}\, d^2B_1\int_\mathcal{S}\, d^2B_2\frac{d\sigma_{A_1A_2\to J/\psi X}}{dyd^2p_\bot d^2B_1d^2B_2}}
{A_1\,A_2\,\frac{d\sigma_{pp\to J/\psi X}}{dyd^2p_\bot }}\,.
\eeq
where $\mathcal{S}$ stands for the overall area of two colliding nuclei. Since the mechanism of $\jpsi$ production in $pp$ collisions remains elusive, we follow our approach in the previous publications and approximate
\beql{prev}
\frac{d\sigma_{pp\to \jpsi X}}{dy}= C\, \frac{d\sigma_{AA\to \jpsi X}}{dy}\bigg|_{A=1}
\eeq
with $C=\text{const}$.  We fix the constant to provide the best description of the $pp$ and $dA$ data. 
We found in the previous paper \cite{Dominguez:2011cy} that is close to unity, so we set $C=1$ in the present paper as well.

\section{Numerical calculations}\label{sec:num}

Experimental data for the differential cross section of $\jpsi$ production in $pp$ collisions can be parameterized in the following form \cite{Adare:2006kf}:
\beql{Fpp}
\frac{d\sigma_{pp\to \jpsi X}}{\sigma_{pp}\, d^2 p_\bot}= \mathcal{N} \left(1+\frac{p_\bot^2}{p_0^2}\right)^{-6}
\eeq
where $\mathcal{N}$  and $p_0$ are positive constants and $\sigma_{pp}$ is the total inelastic cross section.
From \eq{F-four} it follows that 
\beql{intF}
F(\Delta)= \frac{(p_0\Delta)^5}{384}K_5(p_0\Delta)\,.
\eeq
We calculate the spectrum in $pp\to \jpsi X$ process by setting $A_1=A_2=1$ in \eq{xsec3}. Although the saturation momenta are small in this case, there is still a small influence of higher twist terms on the final spectrum which results in a shift of the average transverse momentum, which according to \eq{Fpp} equals $\aver{p_\bot^2}= p_0^2/4$. Because of this we cannot directly use the experimentally determined value of $p_0$, but rather fit the constants $p_0$ and $\mathcal{N}$ to the data \cite{Adare:2006kf,Aamodt:2011gj}\footnote{Since we are concerned with low $p_T$ $\jpsi$'s we  fit $p_0$ and $\mathcal{N}$ to the ALICE collaboration data \cite{Aamodt:2011gj}.}.  

\begin{figure}[ht]
\begin{tabular}{cc}
      \includegraphics[height=4cm]{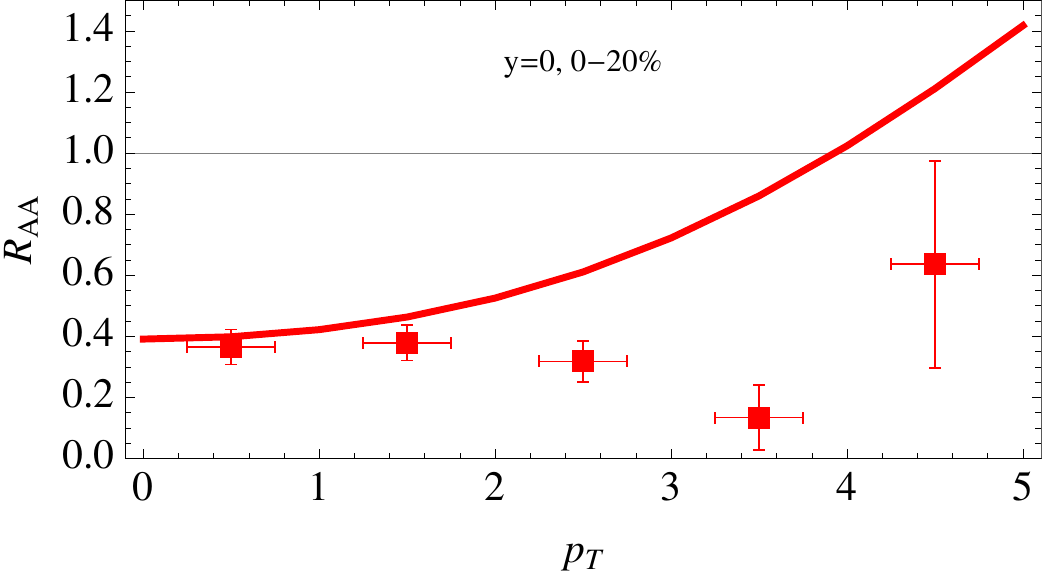} &
      \includegraphics[height=4cm]{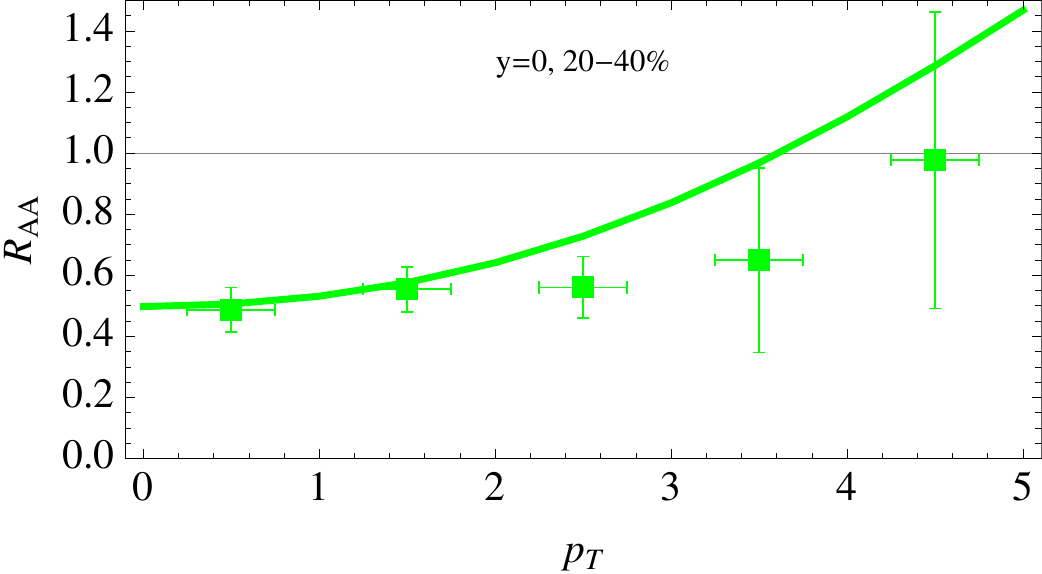}\\
      $(a)$ & $(b)$ \\
      \includegraphics[height=4cm]{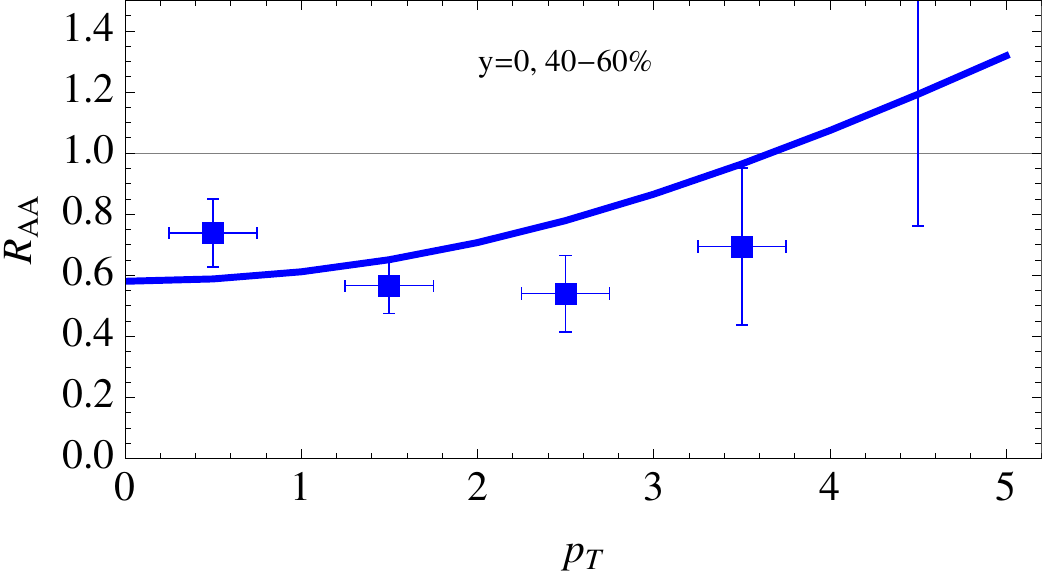} &
      \includegraphics[height=4cm]{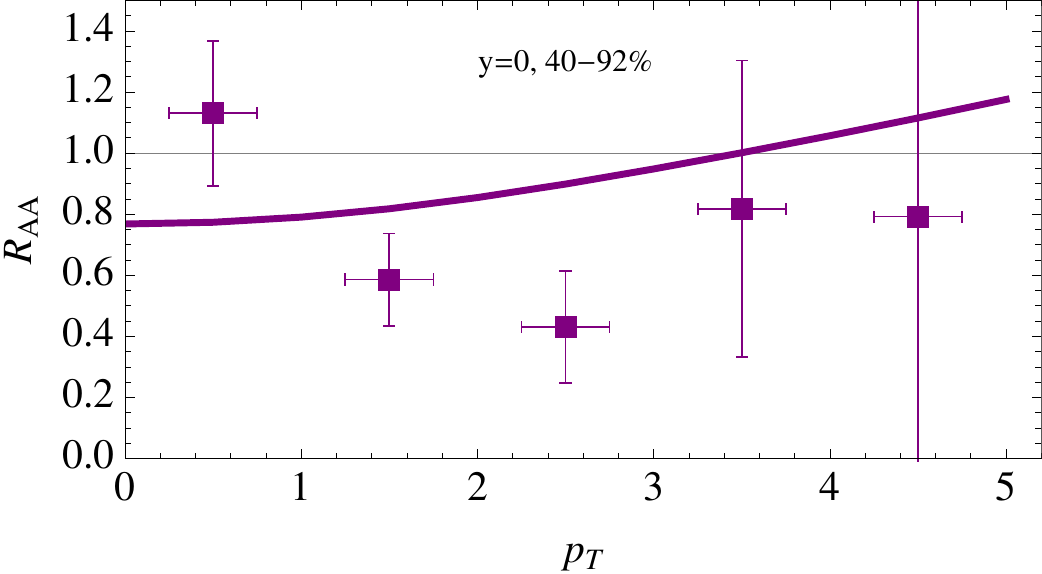}\\
      $(c)$ & $(d)$ \\
       \includegraphics[height=4cm]{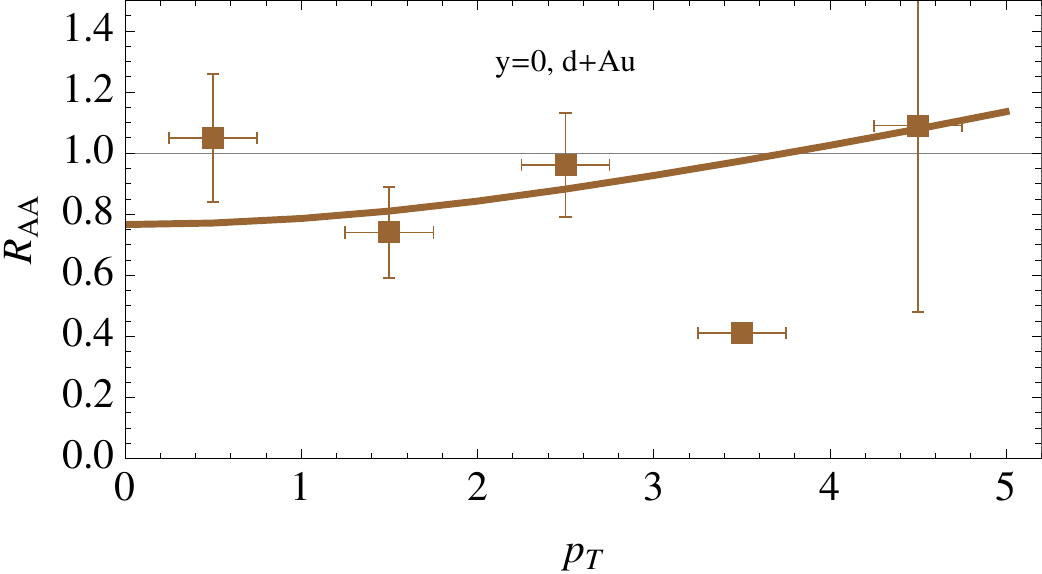}& \\
       $(e)$
      \end{tabular}
  \caption{(Color online). Nuclear modification factor for $\jpsi$'s vs $p_\bot$ in GeV at $\sqrt{s}=200$~GeV, $y=0$ in $AuAu$  for centralities (a)  0-20\%, (b) 20-40\%, (c) 40-60\%, (d) 60-92\% and in (e) minbias $dAu$. Data is from  \cite{Adare:2006ns,Adare:2007gn}.}
\label{raa0}
\end{figure}

We performed  the numerical calculations  using the DHJ model of the dipole scattering amplitude \cite{Dumitru:2005kb}.  The DHJ model is an improvement of the  KKT model \cite{Kharzeev:2004yx,Tuchin:2007pf} that takes into account the change in the anomalous dimension of the gluon distribution function due to the presence of the saturation boundary \cite{Mueller:2002zm} and takes into account some higher order effect. It successfully describes the single inclusive hadron production in $dA$ collisions in the relevant kinematic region.  In this model, the dipole scattering amplitude is parameterized as follows
\beq\label{modN}
N_A(\b r,0, y)=1-\exp\left\{ -\frac{1}{4}\left(r^2 Q_s^2\right)^{\gamma}\right\}\,.
\eeq
 The \emph{gluon} saturation scale  is given by
\beq\label{satt}
Q_s^2=\Lambda^2\, A^{1/3}\, e^{\lambda y}=0.13\,\mathrm{GeV}^2\,e^{\lambda y}\,N_\mathrm{coll}\,.
\eeq
where the parameters $\Lambda=0.6$ GeV and $\lambda=0.3$ are fixed to the low-$x$ DIS data \cite{MOD} and are consistent  \cite{Kharzeev:2000ph,Dumitru:2011wq} with the RHIC and LHC results on hadron multiplicities. The anomalous dimension reads
\begin{align}
\gamma= \gamma_s + (1-\gamma_s)\frac{\ln(M_\bot^2/Q_s^2)}{\lambda Y+ \ln( M_\bot^2/Q_s^2)+d \sqrt{Y} }
\end{align}
 where $M_\bot=\sqrt{p_\bot^2+4m^2}$, $\gamma_s= 0.628$ is implied by theoretical arguments \cite{Mueller:2002zm} and $d=1.2$ is fixed by fitting to the hadron production data in $dA$ collisions at the RHIC.  $Y= \ln(1/x)$, with $x= me^{-y}/\sqrt{s}$. The quark dipole scattering amplitude is given by 
 \begin{align}
 N_F(\b r, 0, y)= 1-\sqrt{1-N_A(\b r, 0, y)}\,.
 \end{align}

\begin{figure}[ht]
\begin{tabular}{cc}
      \includegraphics[height=4cm]{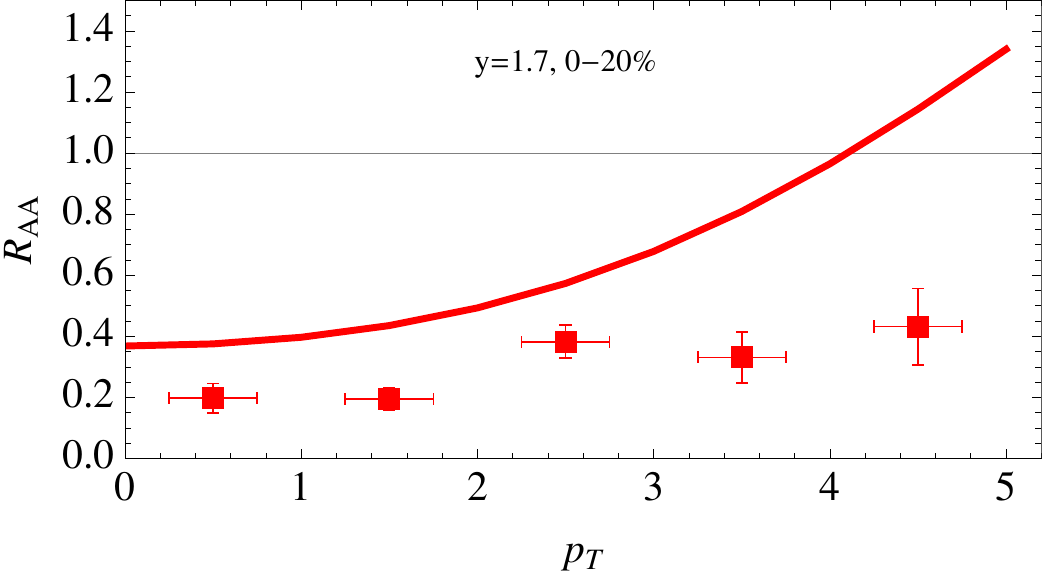} &
      \includegraphics[height=4cm]{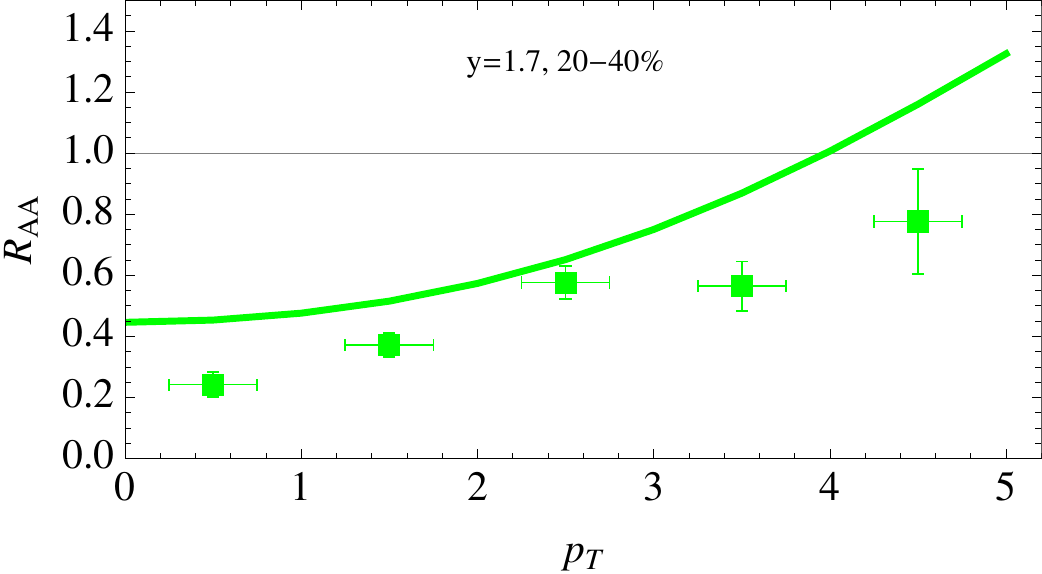}\\
      $(a)$ & $(b)$ \\
      \includegraphics[height=4cm]{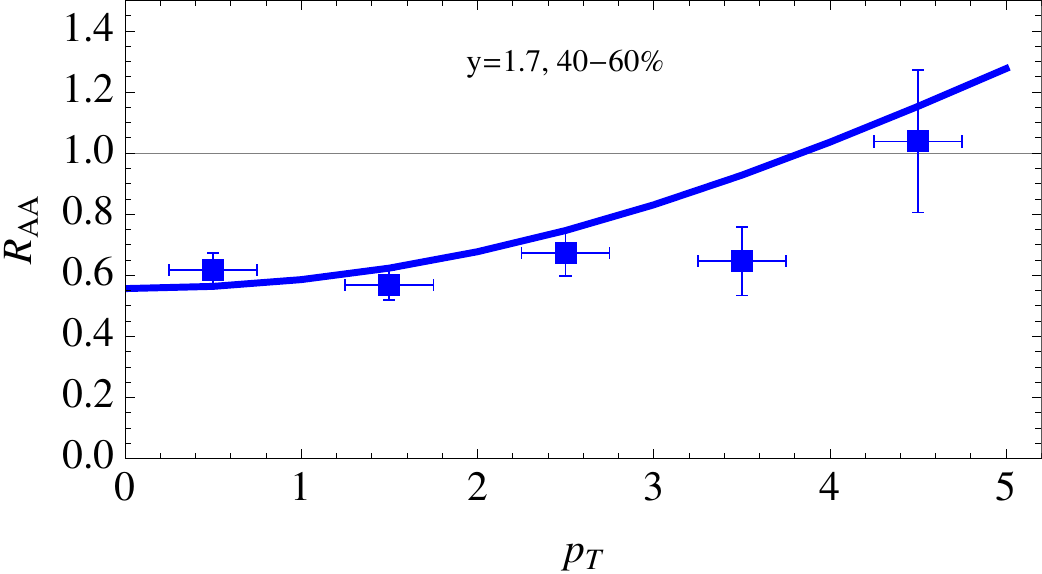} &
      \includegraphics[height=4cm]{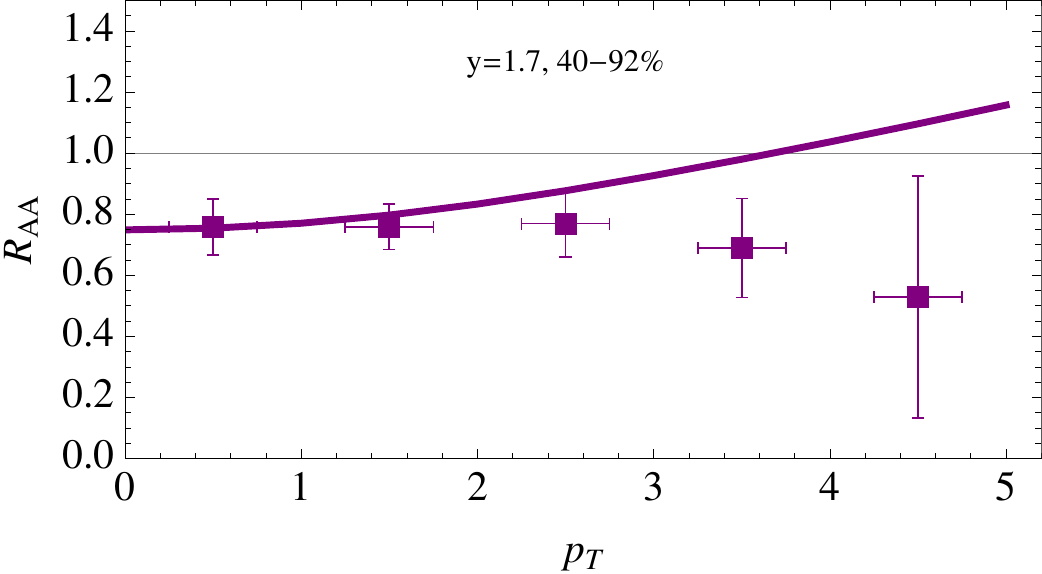}\\
      $(c)$ & $(d)$ \\
       \includegraphics[height=4cm]{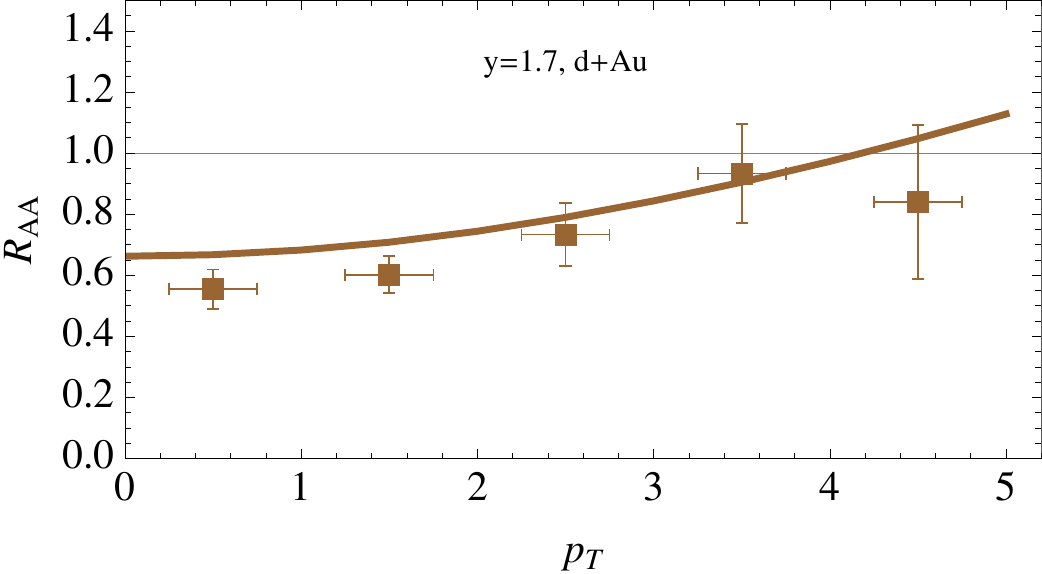}& \\
       $(e)$
      \end{tabular}
  \caption{ (Color online). Nuclear modification factor for $\jpsi$'s vs $p_\bot$ in GeV at $\sqrt{s}=200$~GeV, $y=1.7$ in $AuAu$  for centralities (a)  0-20\%, (b) 20-40\%, (c) 40-60\%, (d) 60-92\% and in (e) minbias $dAu$. Data is from \cite{Adare:2006ns,Adare:2007gn}.}
\label{raaF}
\end{figure}

 In the DHJ model \eq{sec9} reads
 \begin{align}\label{sec10}
\frac{d\sigma_{A_1A_2\to \jpsi X}}{d^2p_\bot dy}=&\frac{d\sigma_{A_1A_2\to \jpsi X}}{ dy }
\int \frac{d^2\Delta}{(2\pi)^2}e^{i\b\Delta\cdot \b p}F(\Delta)  \frac{4}{(Q_{s1}^{2\gamma}+Q_{s2}^{2\gamma})\Delta^{2\gamma}}\left( 
1-e^{-\frac{1}{4}(Q_{s1}^{2\gamma}+Q_{s2}^{2\gamma})\Delta^{2\gamma}}\right)\,,
\end{align}

\begin{figure}[ht]
\begin{tabular}{cc}
      \includegraphics[height=4cm]{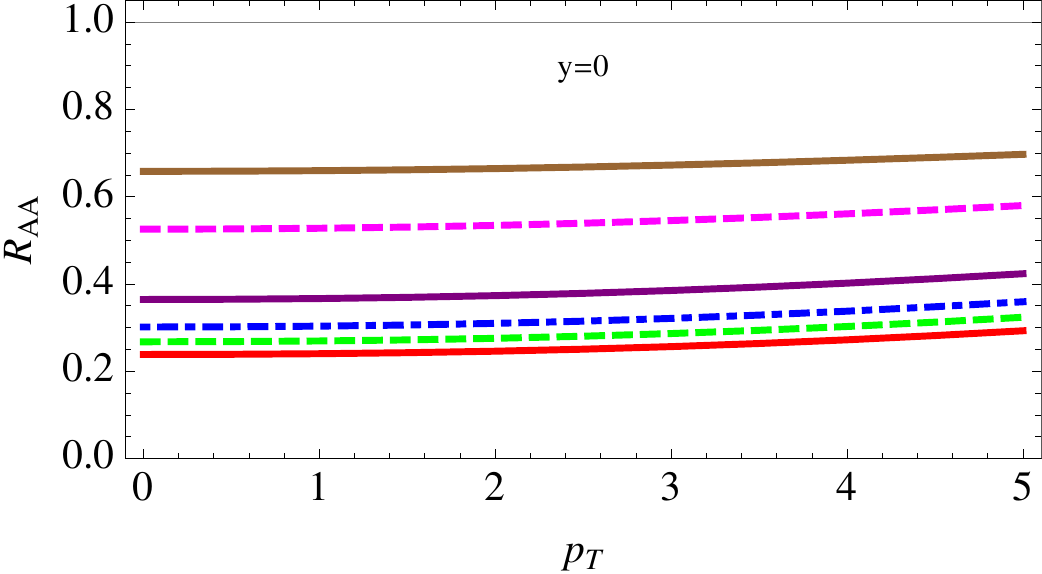} &
      \includegraphics[height=4cm]{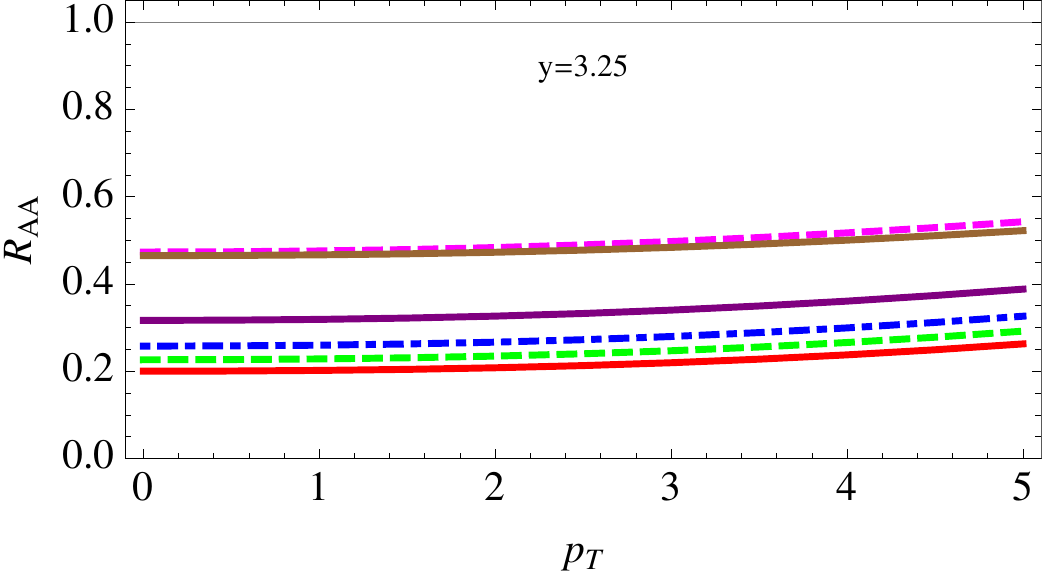}\\
      $(a)$ & $(b)$ 
      \end{tabular}
  \caption{ (Color online). Nuclear modification factor for $\jpsi$'s vs $p_\bot$ in GeV at $\sqrt{s}=7$~TeV in $PbPb$ for rapidities (a) $y=0$ and (b) $y=3.25$. Each line corresponds to a different centrality bin; from bottom to top:  0-10\% (solid red),  10-20\% (dashed green), 20-30\% (dash-dotted blue),  30-50\% (solid purple), 50-80\% (dashed magenta) and in minbias $pPb$ (solid brown).}
\label{raa-lhc}
\end{figure}

The results of our numerical calculations using \eq{sec10} are presented in \fig{raa0},\fig{raaF}  and \fig{raa-lhc} for the center-of-mass energies $\sqrt{s}=200$~GeV and  $\sqrt{s}=7$~TeV. As mentioned in the Introduction they indicate that the final state effects on $\jpsi$ production increase with $p_\bot$. 

\section{Summary}\label{sec:sum}

In this paper we derived the $p_\bot$ spectrum of $\jpsi$'s produced in $pA$ and $AA$ collisions in the framework of the dipole model. We took into account the strong coherence effects in the cold nuclear medium, but entirely neglected the final state effects. We used a phenomenological model for the scattering amplitude to numerically  investigate the transverse momentum dependence of the nuclear modification factors. Our results provide a useful reference for evaluation of the contribution of the final state effects to the $\jpsi$ suppression; they are reasonable agreement with RHIC data on $\jpsi$ production in $dAu$ collisions. The LHC $pA$ data will be an important test of our approach based on the gluon saturation. We find that the $\jpsi$ suppression that originates from the initial state (cold nuclear matter) effects increases at the LHC energies compared to RHIC. Meanwhile, the experimental data on $AA$ collisions indicate \cite{Abelev:2012rv} that $\jpsi$'s are suppressed less at LHC than at RHIC. We have not found a solution for this problem -- in fact, our results exacerbate it.

\acknowledgments
The work of  D.K.\ was supported in part by the U.S.\ Department of Energy under Contracts No. DE-AC02-98CH10886 and DE-FG-88ER41723.
K.T.\ was supported in part by the U.S.\ Department of Energy under Grant No.\ DE-FG02-87ER40371. This research of E.L.\ was supported in part by the Fondecyt (Chile) grant 1100648.

\appendix
\section{The longitudinal form factor $\mathcal{F}_L$}\label{appB}

\begin{figure}[ht]
     \includegraphics[height=3cm]{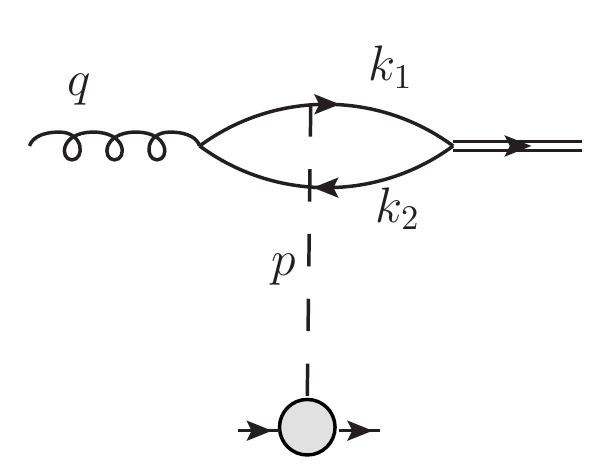}
  \caption{A diagram contributing to the  form factor. Dashed line represents all possible gluon exchanges.}
\label{f:long-ff}
\end{figure}

The longitudinal form factor $\mathcal{F}_L$ is a measure of the coherence of a high energy process. It is a function of the longitudinal momentum transfer $p_z$ and is defined as 
\beql{long-ff}
\mathcal{F}_L(p_z)= \frac{1}{A}\int d^2 b \int_{-\infty}^\infty d\xi\, \rho(\b b, \xi)\, e^{ip_z \xi}
\eeq
where $\xi $ is the longitudinal coordinate. In the coherent regime $p_z \xi \ll 1$ implying that $\mathcal{F}_L=1$. Denote the momenta of gluon, quark and antiquark  as $q=(\omega, \b 0,\omega)$, $k_1=(z\omega,\b k_1,k_{1z})$ and $k_2=((1-z)\omega, \b p -\b k_1,k_{2z})$, see   \fig{f:long-ff}.  The longitudinal momentum transfer $p_z$  can be evaluated as 
\beql{del1}
p_z= q_z-k_{1z}-k_{2z}\approx \frac{\b k_1^2+m^2}{2z\omega}+\frac{(\b p-\b k_1)^2+m^2}{2(1-z)\omega}
\eeq
The relative transverse momentum of $c $ and $\bar c$ is small compared to the typical transverse momentum of the $\jpsi$ center of mass, viz.\  $|\b k_1-\b k_2|\sim \as m\ll p_\bot$. Therefore, we can approximately set $k_{1\bot}\approx k_{2\bot}\approx p_\bot/2$. Considering that typically $z\sim 1/2$ we derive 
\beql{del2}
p_z= \frac{1}{\omega}\left[p_\bot^2+(2m)^2\right] = \frac{2}{\sqrt{s}}e^{-y}\left[p_\bot^2+(2m)^2\right] \,.
\eeq

\begin{figure}[ht]
      \includegraphics[height=4cm]{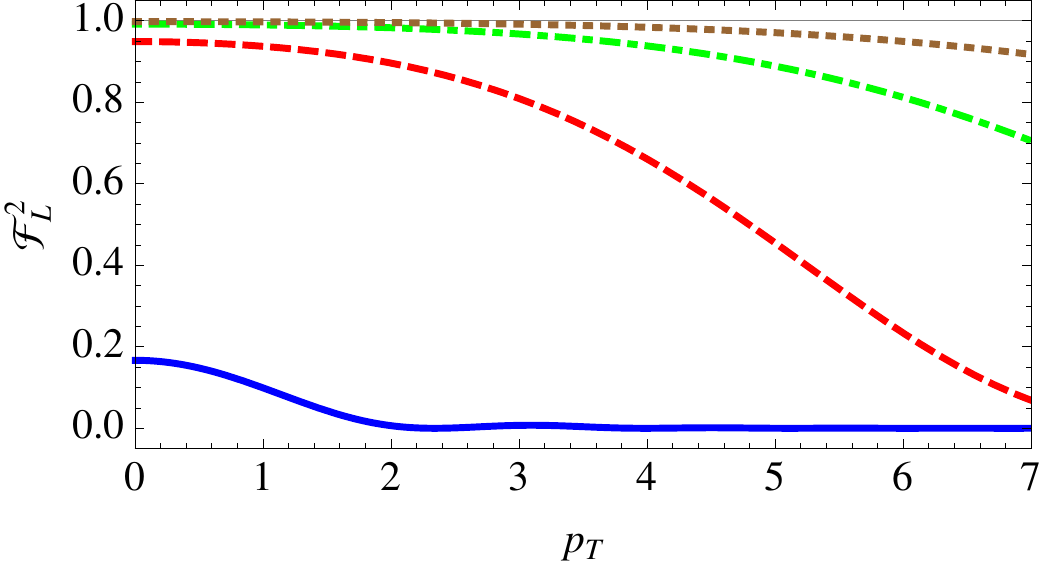}
        \caption{(Color online). Square of the longitudinal form factor as a function of transverse momentum of $\jpsi$ in GeV's.  Lines from bottom to top: $\sqrt{s}=0.2$~TeV, $y=0$ (blue, solid), $\sqrt{s}=0.2$~TeV, $y=1.7$ (red, dashed), $\sqrt{s}=2.76$~TeV, $y=0$ (green, dot-dashed), $\sqrt{s}=5.5$~TeV, $y=0$ (brown dashed).  }
\label{fig:long-ff}
\end{figure}

Consider the ``hard sphere" model of the nucleus:
\beql{hs}
\rho(\b b, \xi) = \frac{A}{\frac{4}{3}\pi R_A^3}\,\theta\left(|\xi|- \sqrt{R_A^2-b^2}\right)\,,
\eeq
where $\theta$ is the step function. Then, substituting \eq{hs},\eq{del2} into \eq{long-ff} we find that  the longitudinal form factor reads
\beql{ff3}
\mathcal{F}_L= \frac{3}{(p_zR_A)^3}\left[ \sin(p_zR_A)-p_zR_A\cos(p_zR_A)\right]\,.
\eeq
The cross section is proportional to $\mathcal{F}_L$ which is plotted in \fig{fig:long-ff}. We see that the dipole model works well at the LHC and at rapidity $y=1.7$, $p_\bot\lesssim 3$~GeV at the RHIC, whereas at $y=0$ and the RHIC it gives a qualitative estimate at best. 

\section{Evaluation of the integrals in \eq{J0}}\label{appA}

Consider a typical  integral appearing in \eq{J0}
\beql{int-1}
I(x)= \int \frac{d^2\ell}{\ell^4}e^{i\b\ell\cdot \b x}
\eeq
This integral is quadratically divergent in the infrared region. However, the quadratic divergent terms cancel   between the four terms appearing in \eq{J0} as is evident in \eq{tot-m-exp}. Therefore, we are interested only in terms that diverge at most logarithmically. To find those,  take Laplacian of $I(x)$
\beql{int-2}
\partial^2_x I(x)  = -\int \frac{d^2\ell}{\ell^2}e^{-i\b\ell\cdot \b x} = -2\pi \ln \frac{1}{x\mu}
\eeq
Note that $I(x)$ depends only on the absolute value of $x$. Using the polar coordinates we cast \eq{int-2} in the form 
\beql{int-3}
\frac{1}{x}\frac{\partial}{\partial x}\left( x\frac{\partial}{\partial x}I\right)= -2\pi \ln \frac{1}{x\mu}
\eeq
Integrating this equation yields
\beql{int-4}
I(x)= -(2\pi)\frac{x^2}{4}\ln \frac{e}{x\mu} + \ldots
\eeq
where elapses indicate the divergent terms independent of $x$.


\end{document}